\documentclass[
twocolumn,
superscriptaddress,
preprintnumbers,
nofootinbib,
amsmath,amssymb,
aps,
]{revtex4-2}

\allowdisplaybreaks
\usepackage{graphicx,xcolor}
\usepackage{dcolumn}
\usepackage{bm}
\usepackage{hhline}
\usepackage{soul}

\begin{document}
	
	\preprint{OUTP-22-11P, MS-TP-22-51,
		CERN-TH-2022-142}
	
	\title{Threshold resummation for the production of four top quarks at the LHC}
	
	\newcommand{\OXaff}{Rudolf Peierls Centre for Theoretical Physics, Clarendon Laboratory, Parks Road,
		University of Oxford, Oxford OX1 3PU, UK}
	
\newcommand{\CEaff}{Theoretical Physics Department, CERN, 1211 Geneva 23, Switzerland}	
	
	\newcommand{\MCaff}{Institute for Theoretical Physics, WWU M\"{u}nster, D-48149 M\"{u}nster, Germany}
	
	\newcommand{\order}[1]{\mathcal{O}\left(#1\right)}
	\newcommand{\as}{\alpha_s}
	\newcommand{\MSbar}{\ensuremath{\overline{\text{MS}}}}
	\newcommand{\ftop}{t\bar{t}t\bar{t}}
	\newcommand{\sqrs}{\sqrt{\hat s}}	
	\newcommand{\rhohat}{\hat \rho}	
	\newcommand{\dd}{\;\mathrm{d}}
	\newcommand{\commentmvb}[1]{\textcolor{blue}{{\bf [#1]$_\text{MvB}$}}}
	\newcommand{\mvb}[1]{\commentmvb{#1}}
	\newcommand{\old}[1]{\textcolor{gray}{#1}}
	\newcommand{\new}[1]{\textcolor{red}{#1}}
	\newcommand{\commentlmv}[1]{\textcolor{magenta}{{\bf [#1]$_\text{LMV}$}}}
	\newcommand{\lmv}[1]{\commentlmv{#1}}
	\newcommand{\commentak}[1]{\textcolor{red}{{\bf [#1]$_\text{AK}$}}}
	\newcommand{\ak}[1]{\commentak{#1}}
	
	\newcommand\Tstrut{\rule{0pt}{2.5ex}}         
		
	\author{Melissa van Beekveld }%
	\email{melissa.vanbeekveld@physics.ox.ac.uk}
	\affiliation{\OXaff}
	\author{Anna Kulesza}%
	\email{anna.kulesza@uni-muenster.de}
	\affiliation{\MCaff} 
	\affiliation{\CEaff}
	\author{Laura Moreno Valero }%
	\email{l\_more02@uni-muenster.de}
	\affiliation{\MCaff}
	
	\date{\today}
	
	\begin{abstract}
		We compute the total cross section for $\ftop$ production at next-to-leading logarithmic (NLL$^{\prime}$) accuracy. This is the first time resummation is performed for a hadron-collider process with four coloured particles in the final state. The calculation is matched to the next-to-leading order strong and electroweak corrections. The  NLL$^{\prime}$ corrections enhance the total production rate by 26\%. The size of the theoretical error due to scale variation is reduced by close to a factor of two, bringing the theoretical error significantly below the current experimental uncertainty of the measurement.
	\end{abstract}
	\maketitle
	The production of four top quarks, $pp \to \ftop$, is one of the rarest Standard Model (SM) production processes currently accessible experimentally at the Large Hadron Collider (LHC). Its cross section is known to receive significant contributions in various SM extensions, hence an accurate measurement can set strong constraints on new physics models. Examples of such scenarios include supersymmetric theories, where the $\ftop$ signal can be enhanced by squark and gluino decays~\cite{FARRAR1978575, Toharia:2005gm}, the production of a new heavy (pseudo)scalar boson in association with a $t\bar{t}$ pair~\cite{Dicus:1994bm,Craig:2015jba,Craig:2016ygr}, or pair production of scalar gluons~\cite{Plehn:2008ae, Calvet:2012rk, Beck:2015cga, Darme:2018dvz}. Moreover, the $\ftop$ production rate is sensitive to the Yukawa coupling of the top quark, making it a useful process to further constrain the nature of Higgs-top quark interactions~\cite{Cao:2016wib, Cao:2019ygh}. When interpreted in the framework of an effective theory, a measurement of the $\ftop$ production process  places strong constraints on the four-fermion operator~\cite{Zhang:2017mls, Banelli:2020iau,Aguilar-Saavedra:2018ksv,Hartland:2019bjb,Ethier:2021bye, Darme:2021gtt, Aoude:2022deh}.

	The ATLAS and CMS experiments have searched for the production of $\ftop$ at the LHC operating at $\sqrt{s} = 13$~TeV~\cite{CMS:2019rvj,CMS:2019jsc, ATLAS:2018kxv, ATLAS:2020hpj, ATLAS:2021kqb, CMS-PAS-TOP-21-005}.  In the latest ATLAS analysis~\cite{ATLAS:2021kqb} a cross section of $\sigma_{\ftop} = 24 \pm 4 ({\rm stat.) ^{+5}_{-4} (\rm syst.})$~fb is measured,
	whereas the recent combined analysis of CMS~\cite{CMS-PAS-TOP-21-005} reports a cross section of $\sigma_{\ftop} = 17^{+5}_{-5}$~fb.
   Intriguingly, these values lie above the SM prediction, which is calculated at the next-to-leading order (NLO) accuracy both in the strong (QCD) and electroweak (EW) coupling~\cite{Bevilacqua:2012em, Alwall:2014hca,Maltoni:2015ena,Frederix:2017wme, Jezo:2021smh}, with the ATLAS measurement consistent with the SM result only within $2\sigma$. The NLO calculations carry a theoretical error due to scale variation of around 25\%, which is comparable with the size of the individual errors of the latest ATLAS and CMS measurements. It is therefore of crucial importance to improve the precision of the theoretical predictions for the $\ftop$ production, especially having in mind thatfuture analyses involve much larger sets of LHC data and the precision of the measurement will increase substantially.
	
More than 90\% of the full NLO result originates from pure QCD interactions. Currently, the calculation of the next-to-next-to-leading order QCD corrections remains out of reach.  However, it is possible to systematically consider a part of  higher-order QCD corrections originating from multiple soft-gluon emissions.  Given the very large partonic centre-of-mass (CM) energy  $\sqrs$  needed to produce  four top quarks, $\sqrs \gtrsim 700$ GeV, the $\ftop$ production at the LHC very often takes place close to production threshold, with any additional real radiation strongly suppressed. One can therefore expect that a large part of the higher-order corrections is due to soft emission and stems from the threshold region. Correspondingly, computing higher-order corrections of this type offers a promising way to improve the precision of the prediction.   

Higher-order QCD corrections from soft gluon emission can be accounted for using resummation, either in direct QCD or in the soft-collinear effective-field-theory framework. The resummation programme for processes involving multiple top quarks has been very successful over the recent years, leading to substantial improvements of theoretical precision for the calculation of the total production cross section for such processes, such as top-pair production~\cite{Kidonakis:1997gm,Contopanagos:1996nh,Bonciani:1998vc,Kidonakis:2001nj, Czakon:2009zw,Beneke:2009rj,Ahrens:2010zv,Cacciari:2011hy,Beneke:2011mq,Czakon:2018nun} or  $t\bar{t}H/Z/W^{\pm}/\gamma$~\cite{Kulesza:2015vda, Kulesza:2016vnq, Kulesza:2017ukk, Kulesza:2018tqz, Kulesza:2020nfh, vanBeekveld:2020cat, Li:2014ula, Broggio:2015lya, Broggio:2016zgg, Broggio:2016lfj, Broggio:2017kzi, Broggio:2019ewu,Kidonakis:2022qvz}. However, in contrast to $\ftop$, these processes involve at most two coloured particles in the final state.  To the best of our knowledge, resummation for processes involving a higher number of coloured particles has not been achieved before.~\footnote{The calculation of the one-loop soft-anomalous dimension, needed for an NLL($^\prime$) resummation, is performed for processes with $3$ coloured final-state particles, e.g.~\cite{Kyrieleis:2005dt,Sjodahl:2009wx, Chargeishvili:2022ngl}.} 

In this work, we perform for the first time the resummation of a process with $4$ final-state coloured particles at the Born level by applying direct QCD resummation methods in Mellin space to the process $pp \to \ftop$. The calculations are carried out at the next-to-leading logarithmic (NLL) accuracy, and take into account constant $\mathcal{O}(\as)$ non-logarithmic contributions that do not vanish at threshold (leading to NLL$^\prime$ accuracy). 

\section{Methodology}
\label{sec:theory}
Soft-gluon corrections get large at the absolute production threshold when $\sqrs$ approaches $M \equiv 4 m_t$, with $m_t$ the top-quark mass. This corresponds to the limit  $\rhohat \to 1$ with $\rhohat \equiv M^2/\hat s$, and logarithmic behaviour $\alpha_s^n \ln(1-\hat\rho)^{2n}$. The theory of $2\to 4$ threshold resummation builds on that for $2\to 2$ processes~\cite{Kidonakis:1998bk, Contopanagos:1996nh, Kidonakis:1998nf,Bonciani:2003nt}. We work in Mellin space, where the hadronic cross section $\sigma_{\ftop}(N)$ is the Mellin transform w.r.t.~the variable $\rho \equiv M^2/s $
\begin{align}
	\label{eq:mellin-space}
	\sigma_{\ftop}(N) = \int_0^1 {\rm d}\rho \,\rho^{N-1}\sigma_{\ftop}(\rho)\,
\end{align}
of the hadronic cross section in momentum space
\begin{align}
\label{eq:normal-space}
\sigma_{\ftop}(\rho)  = \sum_{i,j} & \int_0^1 {\rm d}x_1 f_i(x_1,\mu_F^2)\int_0^1 {\rm d}x_2 f_j(x_2,\mu_F^2) \nonumber\\
& \times \int_{\rho}^1{\rm d}\rhohat \, \delta\left(\rhohat-\frac{\rho}{x_1x_2}\right)\,\hat{\sigma}_{ij\to \ftop}(\rhohat)\,. 
\end{align}
We use $f_i$ to denote parton distribution functions (PDFs), $\mu_F$ the factorisation scale, and $x_{1,2}$ the momentum fraction of the two colliding partons $i,j$. 
Two partonic channels contribute at leading order (LO), $ij=\{ q\bar q ,gg\}$. The cross section $\hat{\sigma}_{ij\to \ftop}(N)$ is a perturbative function that obeys a refactorisation in the soft and collinear limits into functions containing information on particular modes of dynamics. Correspondingly, one can identify a soft function $\mathbf{S}$, containing corrections originating from soft gluon radiation, a collinear(jet) function for each initial-state leg $\Delta_i$, containing corrections from collinear  gluon radiation. All terms that are non-logarithmic in the soft-gluon limit reside in the hard function $\mathbf{H}$. These functions are defined at the cross section level, i.e.~they include the necessary phase-space integrals. The refactorisation in Mellin space takes the form 
\begin{align}
	\label{eq:resummationdef}
	\hat{\sigma}^{\rm res}_{ij\to \ftop}(N) = \,\,&\Delta_i(N+1) \Delta_j(N+1) \\
	&\hspace{0.3cm} \times \, {\rm Tr} \left[ \mathbf{\bar{S}}_{ij\to \ftop}(N+1) \otimes  \mathbf{H}_{ij\to \ftop}(N)\right]\,, \nonumber
\end{align}
suppressing the dependence on $\mu_R$ and $\mu_F$. 
Jet and soft functions both capture soft-collinear enhancements, so one must subtract the overlap contributions through dividing out the eikonal jet functions $\mathcal{J}_i$ from the soft function. This results in a new soft-collinear-subtracted soft function denoted by $\mathbf{\bar{S}}$, and related to the full one as
\begin{align}
\mathbf{\bar{S}}(N+1) =	\frac{\mathbf{S}(N+1)}{\mathcal{J}_1(N+1) \mathcal{J}_2(N+1) }\,.
\end{align}	
The soft and hard functions are matrices in colour space, indicated by their bold font, and colour-connected, indicated by the $\otimes$-symbol. We now go over the definition of the ingredients in Eq.~\eqref{eq:resummationdef}.

The hard function $\mathbf{H}_{ij\to \ftop}$ obeys the perturbative expansion
\begin{eqnarray}
	\mathbf{H}_{ij\to \ftop} = \mathbf{H}_{ij\to \ftop}^{(0)} + \frac{\as}{\pi}\mathbf{H}_{ij\to \ftop}^{(1)} + \mathcal{O}(\alpha_s^2)\,.\,\,\,\,\,\,
\end{eqnarray}
At the NLL accuracy we need $\mathbf{H}_{ij\to \ftop}^{(0)}$, defined as a matrix in colour space with an element $IJ$
\begin{align}
	\label{eq:hard-func-mellin}
\mathbf{H}_{ij\to \ftop,IJ}^{(0)} = \,\,&\frac{1}{2\hat{s}} \int_0^1{\rm d}\hat{\rho} \, \hat{\rho}^{N-1} \int {\rm d}\Phi^B \sum_{{\rm colour, spin}} \mathcal{A}^{(0)}_I  \mathcal{A}^{\dagger(0)}_J\,, 
\end{align}
where we sum (average) over final(initial)-state colour and polarisation degrees of freedom. 
The Born phase space is denoted by $\Phi^B$. The object $\mathcal{A}^{(0)}_I=\langle c_I |\mathcal{A}^{(0)}\rangle$ is the colour-stripped amplitude projected to the colour-vector $c_I$, with $|\mathcal{A}^{(0)}\rangle$ the amplitude in the corresponding colour basis, and $\mathcal{A}^{(0)\dagger}_J$ is its complex conjugate. 
The full $N$ dependence of the hard function is kept.
We obtain the squared matrix elements numerically from aMC@NLO~\cite{Frederix:2018nkq,Alwall:2014hca}.

The coefficient $\mathbf{H}_{ij\to \ftop}^{(1)}$ enters formally at next-to-next-to-leading logarithmic accuracy but can be used to supplement the NLL expressions, resulting in NLL$^\prime$ precision. It consists of virtual one-loop corrections, $\mathbf{V}_{ij\to \ftop}^{(1)}$, and constant terms stemming from collinear-enhanced contributions $\mathbf{C}_{ij\to \ftop}^{(1)}$ that are not yet captured by the initial-state jet functions $\Delta_i$, i.e.
\begin{align}
	\label{eq:h1}
	\mathbf{H}_{ij\to \ftop}^{(1)} = \mathbf{V}_{ij\to \ftop}^{(1)} + \mathbf{C}_{ij\to \ftop}^{(1)}\,.
\end{align}
While the $\mathbf{C}_{ij\to \ftop}^{(1)}$ coefficient is calculated analytically, $\mathbf{V}_{ij\to \ftop}^{(1)}$ is obtained numerically using MadLoop~\cite{Hirschi:2011pa, Ossola:2007ax, Cascioli:2011va, Denner:2016kdg}. The infrared pole structure of the MadLoop calculation, using FKS subtraction~\cite{Frixione:1995ms, Frixione:1997np, Frederix:2009yq}, matches that of our resummed calculation. 

For the incoming jet functions we use the well-known expressions that can be found in e.g.~\cite{Catani:1989ne, Catani:1998tm, Catani:2003zt}, which are a function of $\lambda = \alpha_s b_0 \ln \bar{N}$ with  $\bar{N} \equiv N{\rm e}^{\gamma_E}$.
The soft function is given by~\cite{Contopanagos:1996nh, Kidonakis:1998bk}
\begin{align}
	\label{eq:softdefinition}
	\mathbf{S}_{ij\to \ftop}= \mathbf{\overline{U}}_{ij\to \ftop}\,\,\mathbf{\widetilde{S}}_{ij\to \ftop}\,\,\mathbf{U}_{ij\to \ftop} \,,
\end{align}
with the evolution matrix written as a path-ordered exponential
\begin{align}
	\label{eq:uevol}
	\mathbf{U}_{ij\to \ftop} = P {\rm exp}\left[\frac{1}{2}\int_{\mu_R^2}^{M^2/\bar{N}^2}\frac{{\rm d}q^2}{q^2} \mathbf{\Gamma}_{ij\to \ftop}(\alpha_s(q^2))\right],
\end{align}
and $\mathbf{\Gamma}_{ij\to \ftop}(\alpha_s(q^2))$ the soft anomalous dimension (AD) matrix. 
To achieve NLL($^\prime$) resummation one needs the one-loop contribution $\mathbf{\Gamma}_{ij\to \ftop}^{(1)}$ in Eq.~\eqref{eq:uevol}. This object consists of a kinematic part and a colour-mixing part, which accounts for the change in colour of the hard system, i.e.
\begin{eqnarray}
	\label{eq:sad}
\mathbf{\Gamma}_{ij\to \ftop, IJ}^{(1)} =  \sum_{k,l=1}^6 {\rm Tr}\left[c_I \mathbf{T}_k \cdot \mathbf{T}_l c^{\dagger}_J \right]\Gamma_{kl}\,,
\end{eqnarray}
where $\mathbf{T}_k$ are colour operators. The explicit expression for $\mathbf{\Gamma}_{ij\to \ftop, IJ}^{(1)}$ depends on a choice of basis tensors represented by $c_{I}$ (and $c_J^{\dagger}$ for the complex conjugate) for the underlying hard scattering process $ij \to \ftop$. The kinematic part, $\Gamma_{kl}$, is given by the residue of the UV-divergent part of the one-loop eikonal contributions~\cite{BOTTS198962, Kidonakis:1997gm, Kidonakis:1996aq}. 

The matrix $\mathbf{\widetilde{S}}_{ij\to \ftop}$ in Eq.~\eqref{eq:softdefinition} represents the boundary condition for the solution of the renormalisation group equation at $\mu_R = M/\bar{N}$ from which Eq.~\eqref{eq:softdefinition} follows. Like $\mathbf{H}$, it obeys a perturbative expansion
\begin{align}
	\mathbf{\widetilde{S}}_{ij\to \ftop}  =\mathbf{\widetilde{S}}_{ij\to \ftop}^{(0)} + \frac{\alpha_s}{\pi} \mathbf{\widetilde{S}}_{ij\to \ftop}^{(1)} + \mathcal{O}(\alpha_s^2)\,.
\end{align}
The lowest-order contribution $\mathbf{\widetilde{S}}_{ij\to \ftop}^{(0)}$ is given by the trace of the colour basis vectors for the underlying hard process. For NLL$^\prime$ resummation we also need the first-order correction $\mathbf{\widetilde{S}}_{ij\to \ftop}^{(1)}$, which is calculated analytically by considering the eikonal corrections to $\mathbf{\widetilde{S}}_{ij\to \ftop}^{(0)}$. 

The major difficulty in the resummed calculations for the $\ftop$ production cross section stems from the complicated colour structure of the underlying hard process, involving six coloured particles.  The colour structure of the $q\bar{q} \to \ftop $ process is
\begin{eqnarray}
	\mathbf{3}\otimes \mathbf{\bar{3}} &=& 
	\mathbf{3}\otimes \mathbf{\bar{3}} \otimes 	\mathbf{3}\otimes \mathbf{\bar{3}}.
\end{eqnarray}
The decomposition into irreducible representations reads
\begin{eqnarray}
	\label{eq:reductionqqbar}
	 \mathbf{1}\oplus\mathbf{8} = 
	(2\times\mathbf{1}) \oplus (2\times\mathbf{8}) \oplus \mathbf{8}_S \oplus \mathbf{8}_A \oplus \mathbf{10} \oplus \mathbf{\overline{10}}\oplus \mathbf{27}\,. \,\,\,\,\,\,\,\,\,\,\,
\end{eqnarray}
For the $gg$ channel we have 
\begin{eqnarray}
	\mathbf{8}\otimes \mathbf{8} &=& 
	\mathbf{3}\otimes \mathbf{\bar{3}} \otimes 	\mathbf{3}\otimes \mathbf{\bar{3}} \,,
\end{eqnarray}
and in terms of irreducible representations
\begin{eqnarray}
\mathbf{0}\oplus \mathbf{1}\oplus \mathbf{8}_S \oplus \mathbf{8}_A \oplus \mathbf{10} \oplus \mathbf{\overline{10}}\oplus \mathbf{27} &=& \\
&&\hspace{-4.5cm}	\mathbf{0}\oplus (2\times\mathbf{1}) \oplus (2\times\mathbf{8}) \oplus \mathbf{8}_S \oplus \mathbf{8}_A \oplus \mathbf{10} \oplus \mathbf{\overline{10}}\oplus \mathbf{27}\,. \nonumber
\end{eqnarray}
From this we infer that the $q\bar{q}$ colour space is $6$-dimensional, whereas the $gg$ one is $14$-dimensional.

The one-loop soft AD matrices $\mathbf{\Gamma}_{ij\to \ftop}^{(1)}$ are in general not diagonal. Solving Eq.~\eqref{eq:uevol} in terms of standard exponential functions requires changing the colour bases to $R$ where $\mathbf{\Gamma}_{ij\to \ftop,R}^{(1)}$ is diagonal~\cite{Kidonakis:1998bk}. We find such bases using the technique outlined in Ref.~\cite{Keppeler:2012ih}. The resulting one-loop soft AD matrices for $N_c = 3$ in the threshold limit become~\footnote{Their full forms are provided in the suplementary material~\cite{supplement}.}
\begin{subequations}
\begin{align}
	&	2{\rm Re}[\overline{\mathbf{\Gamma}}_{q\bar{q}\to \ftop,R}]
	= \text{diag}\left(0,0,-3,-3,-3,-3\right), \\
	& 2{\rm Re}[\overline{\mathbf{\Gamma}}_{gg\to \ftop,R}] = \text{diag}(-8, -6, -6, -4, -3,-3,\\
	& \hspace{3cm} -3,-3,-3,-3,-3,-3,0,0). \nonumber
\end{align}
\end{subequations}
The values above are the negative values of the quadratic Casimir invariants for the irreducible representations in which the colour structure of the final state can be decomposed in SU(3). This corresponds to a physical picture where the soft gluon is only sensitive to the total colour charge of a system at threshold, and constitutes a strong check of our calculations. We have verified that the virtual corrections obtained from MadLoop, rewritten in the new basis $R$, are consistently $0$ for the base vector corresponding to a representation whose dimension is zero for $N_c=3$, which is another consistency check of our work. 

With this, the contribution of the soft-collinear-subtracted soft function reads
\begin{align}
	\mathbf{\bar{S}}_{ij\to \ftop, R} (N) &= \\
	&\hspace{-0.2cm}	\mathbf{\bar{\widetilde{S}}}_{ij\to \ftop, R}\, \exp\left[\frac{{\rm Re}[\mathbf{\bar{\Gamma}}_{ij\to \ftop, R}^{(1)}]}{b_0 \pi}\ln\left(1-2\lambda\right)\right], \nonumber
\end{align}
where $\mathbf{\bar{\Gamma}}_{ij\to \ftop, R}^{(1)}$ is related to $\mathbf{\Gamma}_{ij\to \ftop, R}^{(1)}$ after subtracting the soft-collinear contributions~\cite{Kidonakis:1998nf}. The hard function in Eq.~\eqref{eq:resummationdef} is written in the colour tensor basis $R$. 

The last step to calculate a physical cross section in momentum space involves taking the inverse Mellin transform of the $N$-space expression
\begin{align}
	\label{Eq:match}
	\sigma^{\rm NLO +res}_{\ftop}(\rho) & = \sigma^{\rm NLO}_{\ftop}(\rho) +\\
	& \sum_{ij} \int_{\mathcal{C}} \frac{{\rm d}N}{2\pi i}\rho^{-N} f_i(N+1,\mu_F^2) f_j(N+1,\mu_F^2) \nonumber\\
	&  \times \left[ \hat{\sigma}_{ij\to \ftop}^{\rm res}(N) 
	-  \hat{\sigma}_{ij\to \ftop}^{\rm res}(N) \Big|_{{\rm NLO}}
	\right]\,, \nonumber
\end{align}
where `res' denotes LL, NLL or NLL$^\prime$. 
To retain the full information from the perturbative calculation, we match the resummed result to the NLO fixed-order cross section $ \sigma^{\rm NLO}$. To avoid double-counting, the expansion of $\hat{\sigma}_{ij\to \ftop}^{\rm res}(N)$ up to $\mathcal{O}(\as^5)$, $\hat{\sigma}_{ij\to \ftop}^{\rm res}(N) \big|_{{\rm NLO}}$, is subtracted. 
The inverse Mellin transform in Eq.~(\ref{Eq:match}) used the Minimal Prescription~\cite{Catani:1996yz} and is evaluated numerically on a contour $\mathcal{C}$ parameterised by $C_{\rm MP}$ and $\phi_{\rm MP}$ as 
\begin{align}
	N = C_{\rm MP} + y {\rm e}^{i\phi_{\rm MP}}\,,
\end{align}
with $y \in [0,\infty)$. We calculated results for various values $C_{\rm MP}$  and $\phi_{\rm MP}$  to verify the independence of the result on the choice of the contour. 
 Inverting the resummed expression from Mellin- to physical-space involves choosing a specific approach. Other methods than the minimal prescription could be used, see e.g.~\cite{Kidonakis:2001nj, Forte:2006mi}. Different methods reorganise subleading terms differently, leading to numerical differences in the predictions and thus an additional source of uncertainty.

\section{Numerical results}
\label{sec:res}
The phenomenological studies reported in this letter are performed using the central member of the LUXqed\_plus\_PDF4LHC15\_nnlo\_100 PDF set~\cite{Manohar:2016nzj, Manohar:2017eqh} for both the pure QCD results and the QCD+EW results. This PDF set  is based on the PDF4LHC15 PDF set~\cite{Butterworth:2015oua, NNPDF:2014otw, Harland-Lang:2014zoa, Dulat:2015mca} and includes the photon content of the proton.  
We use the $\alpha_s$ value corresponding to the PDF set, take the mass of the top quark $m_t = 172.5$~GeV (unless stated otherwise) and choose the central factorisation and renormalisation scale $\mu_{F,0} = \mu_{R,0} = 2m_t$ (as in Ref.~\cite{Maltoni:2015ena}). The theoretical uncertainty is estimated by varying $\mu_R$ and $\mu_F$ using a $7$-point scale variation, i.e.
\begin{align}
	\label{eq:7-point-variations}
	\left( \frac{\mu_R} { \mu_{R, 0}}, \frac{\mu_F}{\mu_{F,0}} \right)_{7-{\rm point}} \in& \{(0.5,0.5),(0.5,1),(1,0.5),  \nonumber \\
	& (1,1),(1,2),(2,1),(2,2)\}\,.
\end{align}
The fixed-order results are obtained using aMC@NLO~\cite{Frederix:2018nkq,Alwall:2014hca}. Since our calculation concerns pure QCD corrections, we present the LO and NLO QCD results for comparison. However, our final resummation-improved cross section incorporates the NLO(QCD+EW) result, where the electroweak corrections are included up to $\mathcal{O}(\alpha^2)$~\cite{Frederix:2017wme}.~\footnote{In the notation of Ref.~\cite{Frederix:2017wme}, we include terms up to (N)LO$_{3}$.} 

\begin{figure}[t]
	\centering
	\includegraphics[page=1, width=0.45\textwidth]{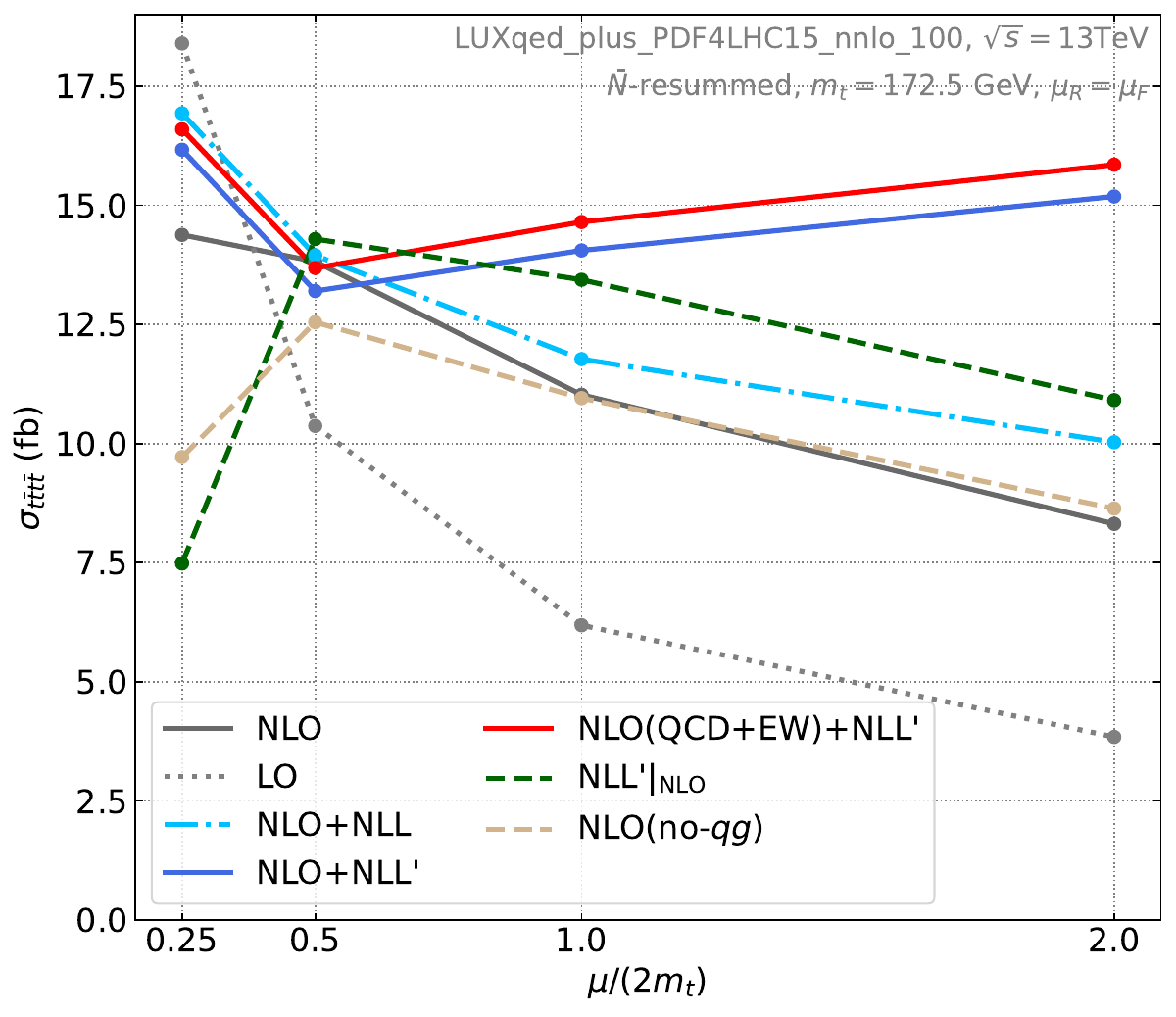}
	\caption{Scale dependence of QCD LO (gray dotted), NLO (gray solid), NLO(no-$qg$) (brown dashed), NLL$^\prime\big|_{\rm NLO}$ (green dashed), NLO+NLL (blue dash-dotted), NLO+NLL$^\prime$ (blue solid) and  NLO(QCD+EW)+NLL$^\prime$ (red solid) cross sections at $\sqrt{s} = 13$~TeV.}
	\label{fig:summary}
\end{figure}

Fig.~\ref{fig:summary} shows the scale dependence of various fixed-order and resummed results for $\sigma_{\ftop}$ setting $\mu_R = \mu_F$. To judge the quality of the resummation we include expanded NLL$^\prime$ resummed result, and the exact NLO cross section without the $qg$ channel, named NLO(no-$qg$).  We observe that apart from the region of very small scales, the expanded result captures more than 90\% of the NLO(no-$qg$) cross section.~\footnote{The difference between the two lines comes for example from the approximation of the soft AD matrix in the absolute-threshold limit, leading to missing kinematic contributions, and not including terms that go as $\mathcal{O}(1/N)$ or higher in the gluon emission contribution.}
While the NLL corrections only moderately improve the scale dependence of the NLO QCD cross section, the scale sensitivity of the NLO+NLL$^\prime$ result is dramatically reduced. NLL$^\prime$ contributions increase the $\sigma_{\ftop}$ predictions by 27\% w.r.t.~the pure NLO QCD result, and by 26\% w.r.t.~the complete NLO(QCD+EW) result, see  Table~\ref{tab:summary}. These corrections are more than three times the size of the previously calculated complete EW effects at NLO. Based on the results presented here we expect that the central value of the NLO+NNLL predictions will be close to the NLO+NLL' result. The former calculation will, however, likely bring down the scale uncertainty of the theoretical predictions, and is left for future work.

\begin{figure}[t]
	\centering
   \includegraphics[page=2, width=0.45\textwidth]{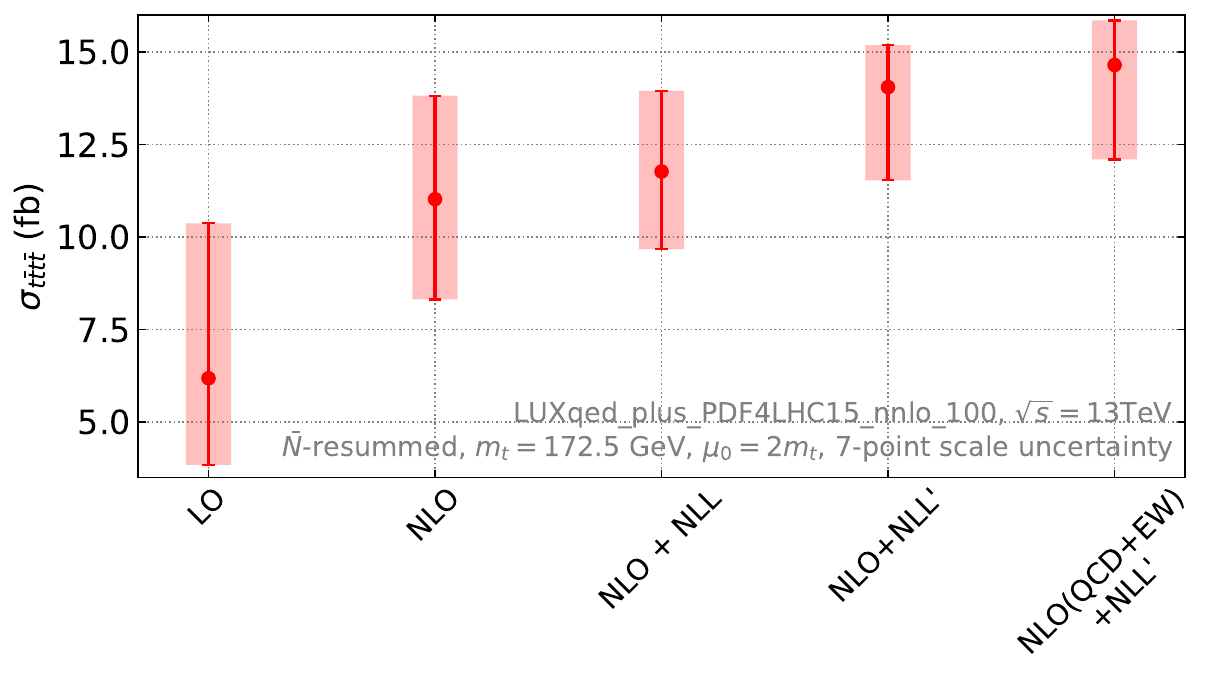}
	\caption{Predictions for the total $pp\rightarrow t\bar{t}t\bar{t}$ cross section at $\sqrt{s}=13$ TeV for fixed-order calculations and resummation-improved results, obtained using Eq.~\eqref{eq:7-point-variations}.}
	\label{fig:summary2}
\end{figure}
\begin{table*}[th]
\begin{tabular}{ccccc}
	$\sqrt{s}$ (TeV) & NLO  & NLO+NLL  &  NLO+NLL$^{\prime}$ &  $ {\rm K_{NLL^{\prime}}}$  
	\\
	\hline \Tstrut
	13 &$11.02(2)_{-24.6\%}^{+25.4\%}$~fb& $11.77(1)_{-17.8\%}^{+ 18.5\%}$~fb& $14.05(1)_{-17.9\%}^{+8.1\%}$ ~fb& $1.27$\\
	13.6 & $13.13(1)_{-24.4\%}^{+25.1\%}$~fb&  $13.99(1)_{-17.8\%}^{+ 18.5\%}$~fb& $16.66(2)_{-18.0\%}^{+8.1\%}$ ~fb& $1.27$\\
	\hline \Tstrut
	$\sqrt{s}$ (TeV) & NLO(QCD+EW)  & NLO(QCD+EW)+NLL  &  NLO(QCD+EW)+NLL$^{\prime}$ &  $ {\rm K_{NLL^{\prime}}}$   \\
	\hline \Tstrut
	13 & $11.62(1)^{+23.1\%}_{-22.7\%}$~fb & 
	$12.37(1)^{+16.7\%}_{-16.4\%}$~fb &  $14.65(2)^{+8.2\%}_{-17.4\%}$~fb & $1.26$\\
	13.6 & $13.83(1)^{+22.6\%}_{-22.9\%}$~fb & 
	$14.69(1)^{+16.7\%}_{-16.4\%}$~fb &  $17.36(2)_{-17.5\%}^{+8.3\%}$~fb & $1.26$\\
\end{tabular}
\caption{\label{tab:summary}
Fixed and resummed-and-matched total cross sections in fb for $pp\to \ftop$ with $\sqrt{s} = 13$~TeV and $\sqrt{s} = 13.6$~TeV, the central scale value of $\mu_0 = 2m_t$ and $m_t = 172.5$~GeV. The number in parenthesis indicates the statistical uncertainty on the last digit whereas the percentage error indicates the $7$-point scale uncertainty (Eq.~\eqref{eq:7-point-variations}). The ${\rm K_{NLL^{\prime}}}$ factor is the ratio of the resummation-improved cross section at NLO+NLL$^\prime$ to the NLO cross section.} 
\end{table*}

Next we examine the reduction of the theoretical error of the resummation-improved cross section using the 7-point method. Table~\ref{tab:summary} summarises the central values of the various predictions together with the corresponding error due to scale variation. This information is graphically represented in Fig.~\ref{fig:summary2}. Remarkably, the scale error of the NLO+NLL$^\prime$ predictions is reduced compared to NLO predictions by close to a factor of two. Including the PDF uncertainty of $\pm 6.9\%$, our state-of-the-art prediction for $\sqrt s=13$ TeV and $m_t = 172.5$~GeV reads
\small
\begin{align}
	\sigma^{\rm NLO(QCD+EW) + NLL'}_{\ftop} =14.65(1) \,^{+1.20}_{-2.55} {\rm (scale)}\pm 1.01 {\rm (pdf)}  {\rm fb}, \nonumber
\end{align}
\normalsize
or, adding the two  theoretical errors in quadrature
	\begin{align}
		\sigma^{\rm NLO(QCD+EW) + NLL'}_{\ftop} =14.65(1) \,^{+1.57}_{-2.75}\  {\rm fb}, \nonumber
	\end{align}
Resumming terms logarithmic in $N$ shifts the central value by 0.26 fb.  We defer a detailed discussion to an upcoming publication~\cite{4top:upcomingpaper}.\footnote{These differences arise from a subset of subleading terms. The treatment of subleading terms will in general differ depending on the approach chosen to invert the Mellin space result. }
In Table~\ref{tab:summary} we also report the obtained cross section for the LHC CM energy of $13.6$~TeV. Including the PDF uncertainty of $\pm 6.7\%$ we obtain 
\small \begin{align}
	\sigma^{\rm NLO(QCD+EW) + NLL'}_{\ftop} & = 17.36(2) \,^{+1.43}_{-3.04} {\rm (scale)}\pm 1.16 {\rm (pdf)}\   {\rm fb} \nonumber \\
	&  = 17.36(2)\,^{+1.85}_{-3.25}\   {\rm fb}, \nonumber 
\end{align} \normalsize
which is an increase of $18.5\%$  w.r.t.~the obtained cross section for $\sqrt{s} = 13$~TeV.

\begin{figure}[t]
	\centering
	\includegraphics[page=1, width=0.45\textwidth]{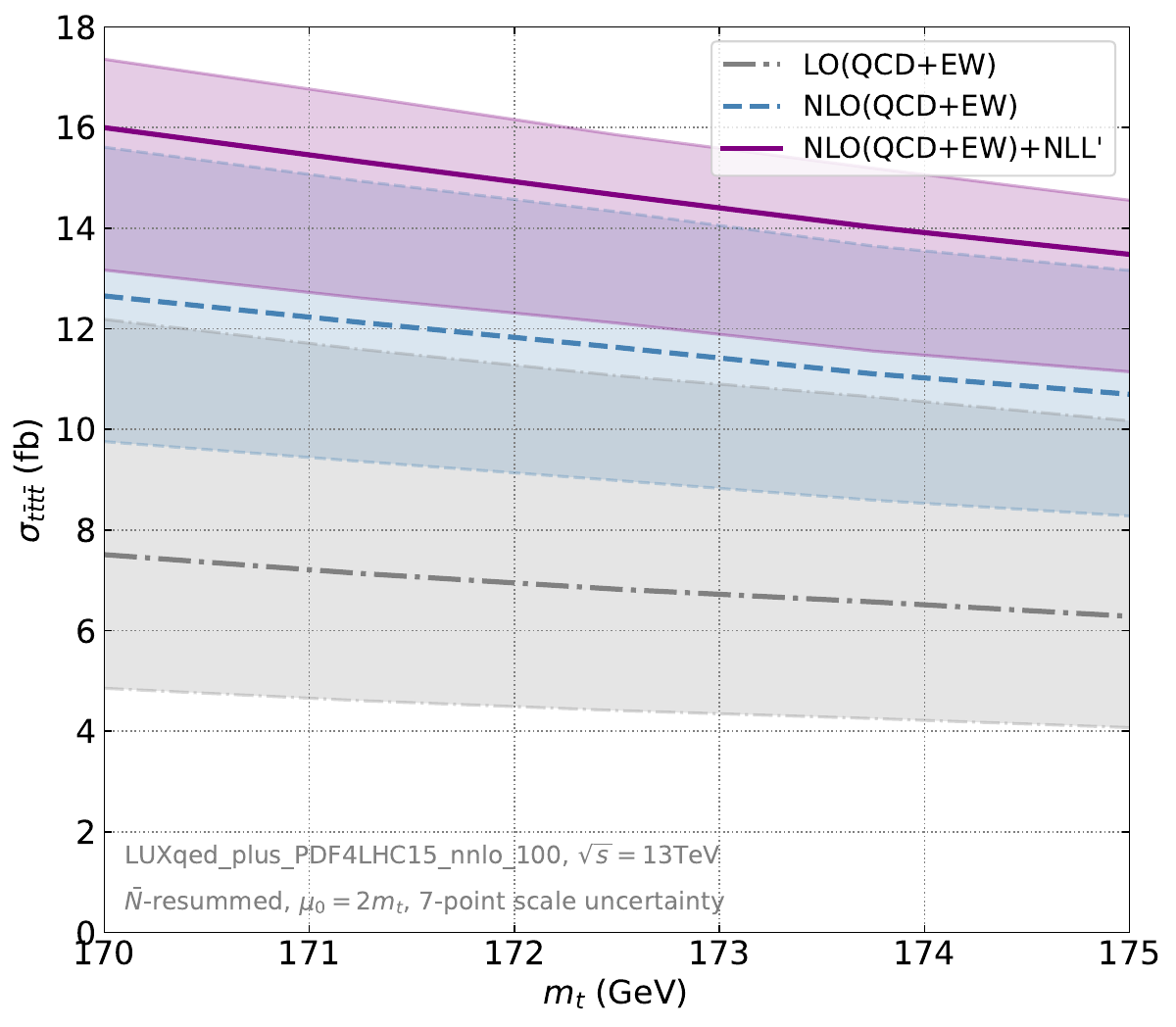}
	\caption{Cross section for the $pp\to \ftop$ process with $\sqrt{s} = 13$~TeV for different values of $m_t$.  Shown are the LO, NLO and NLO+NLL' predictions (QCD + EW). The bands indicates the scale uncertainty calculated using Eq.~\eqref{eq:7-point-variations}.}
	\label{fig:mt-variation}
\end{figure}

We have also studied the effect of varying the value of the top mass in the window of $[170-175]$~GeV. The resulting predictions are shown in Fig.~\ref{fig:mt-variation} for $\sqrt{s} = 13$~TeV. We observe that the correction stemming from soft-gluon resummation is flat under variation of the top quark mass. 

\section{Conclusion}
\label{sec:conclusions}
In this letter, we present predictions for the total cross section of the four top production process at NLO+NLL$^\prime$ accuracy, including electroweak corrections for the fixed-order prediction. This is the first time that the framework of threshold resummation has been applied to a $2\to 4$ process containing six coloured particles at leading order. We present our results both at a collider energy of $13$ and $13.6$ TeV, and vary the top mass in the window of $170-175$ GeV. Setting $m_t = 172.5$~GeV  and $\sqrt{s} = 13.6$~TeV, we find the total cross section $\sigma^{\rm NLO(QCD + EW)+ NLL^\prime}_{\ftop} 17.4^{+8.3\%}_{-17.5\%}$, where the indicated error is estimated using the $7$-point scale uncertainty. When compared to the NLO(QCD+EW)-only prediction, $\sigma^{\rm NLO(QCD+EW)}_{\ftop} = 13.8^{+22.6\%}_{-22.9\%}$~fb, we find that the central value is increased with a $K$-factor of $1.26$. The uncertainty stemming from scale variation is reduced by   close to a factor of two. Including the PDF error in quadrature we reduce the total theoretical uncertainty from $(+23.6\%,-23.9\%)$ at NLO(QCD+EW) to $(+10.7\%,-18.7\%)$ at  NLO(QCD + EW)+ NLL$^\prime$, which lies comfortably below the current experimental uncertainty.
These predictions will play an important role in stress-testing the SM, especially in view of the latest experimental results obtained for $\ftop$ production. 

%
\section*{Acknowlegdements}
We are grateful to Marco Zaro and Davide Pagani for their help in extracting the NLO electroweak corrections from aMC@NLO.  This work has been supported in part by the DFG grant KU 3103/2.
MvB acknowledges support from a Royal Society Research Professorship (RP/R1/180112) and from the Science and Technology Facilities Council under grant ST/T000864/1, while LMV acknowledges support from the DFG Research Training Group “GRK 2149: Strong and Weak Interactions - from Hadrons to Dark Matter". AK gratefully acknowledges the support and the hospitality of the CERN Theoretical Physics Department.

\appendix

\bibliography{MC.bib}

%
%


	\end{document}